\documentclass[aps,prb,twocolumn,amsmath,amssymb,longbibliography,floatfix]{revtex4-2}

\usepackage[utf8]{inputenc}
\usepackage{graphicx}
\usepackage{booktabs}
\usepackage{notes2bib}
\usepackage{amsmath}
\usepackage{siunitx}
\usepackage{soul}

\usepackage[colorlinks=true,bookmarks=false,citecolor=blue,urlcolor=blue,pdfstartview=FitH]{hyperref}

\begin{document}

\title{Complex dispersion relation of Rayleigh-Bloch waves trapped by slow inclusions}
\author{Vincent Laude}
\affiliation{Université Marie et Louis Pasteur, CNRS, institut FEMTO-ST, F-25000 Besançon, France}

\begin{abstract}
Rayleigh-Bloch waves are guided acoustic waves propagating along a periodic line of inclusions placed inside an open, infinite medium. Below the sound cone, they are transversely evanescent on both sides of the line of inclusions. Guidance is then achieved without any cladding surrounding the segmented core. Inclusions usually impose definite boundary conditions, resulting in a single guided band. We consider instead the case of permeable, slow inclusions inside a fast medium. Introducing the concept of guided quasi-normal modes, we obtain the complex dispersion relation taking into account radiation at infinity. We thus show that multiple bands of leaky Rayleigh-Bloch waves appear and that guided bound states in the continuum arise as a result of the combination of symmetry and periodicity.
\end{abstract}

\maketitle


\section{Introduction}
\label{sec1}

\begin{figure}
\includegraphics[width=\columnwidth]{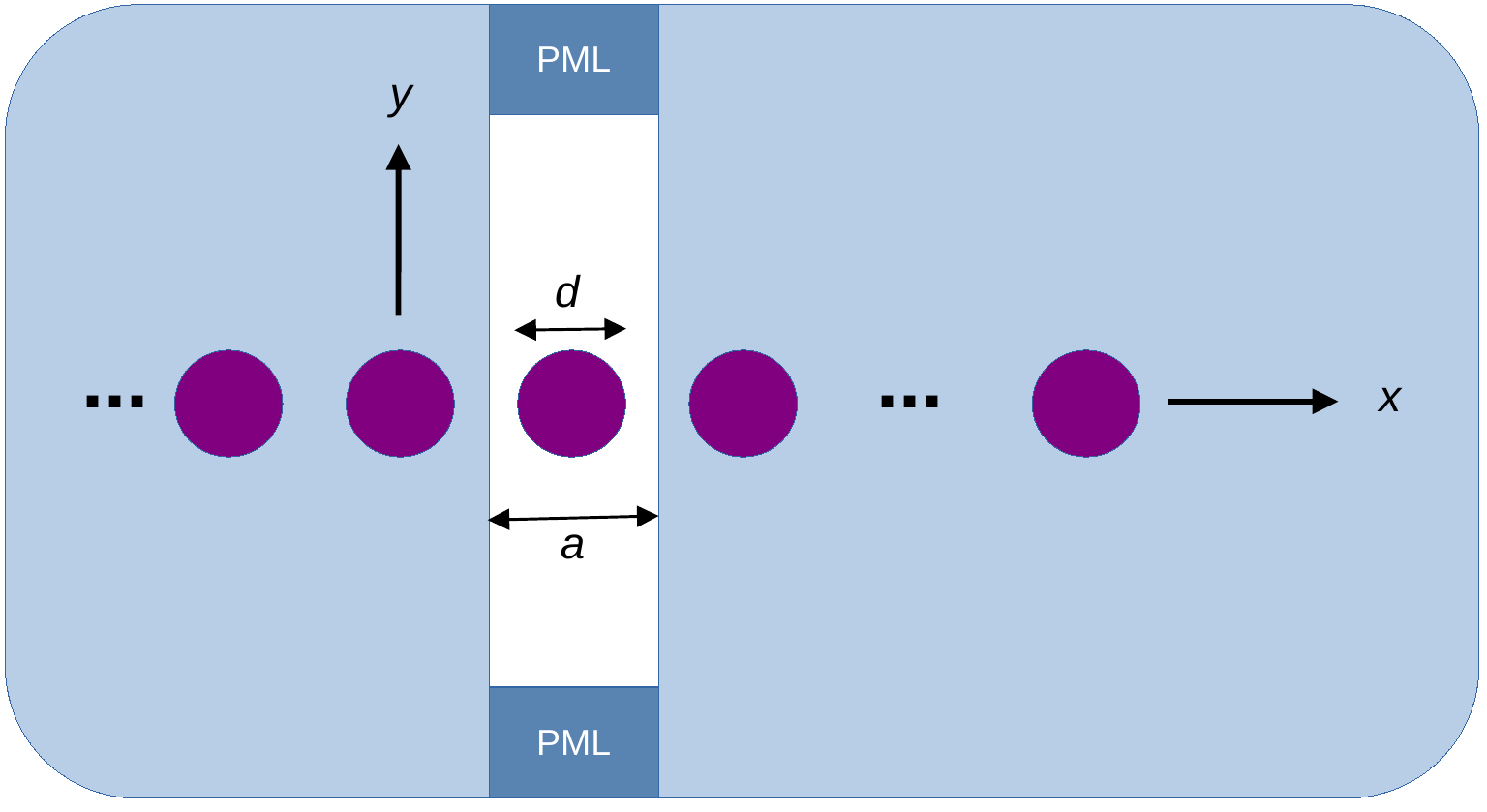}
\caption{Waveguide composed of a line of periodic inclusions in an open, infinite, host medium. The central portion shows the primitive unit-cell with lattice constant $a$, terminated at top and bottom by perfectly matched layers (PML). The diameter of cylindrical inclusions is $d=0.6a$.}
\label{fig1}
\end{figure}

Rayleigh-Bloch (RB) waves are waves guided along a periodic chain of scatterers or inclusions inside an open, infinite, host medium (Figure \ref{fig1}a) \cite{bennetts2022rayleigh}.
The significance of such an arrangement is that the chain acts as a guide and confines waves in space even though there is no physical boundary to contain them.
In essence, the dispersion of Rayleigh-Bloch waves is not determined by boundary conditions but by the periodic distribution of inclusions.
Denoting $x$ the periodic axis (Figure \ref{fig1}), guided modes of propagation are Bloch waves of the form $p(x,y) \exp(\imath (\omega t - k x))$, with $k$ the Bloch wavenumber and $p(x,y)$ a wave field periodic along axis $x$ that decays exponentially in the positive and negative $y$ direction.
Strictly speaking, the dispersion of Rayleigh-Bloch waves lies only under the sound cone, the region of dispersion space for which waves have a phase velocity smaller than that of any bulk wave in the host medium.
Rayleigh-Bloch waves have been discussed for many different physical systems, including  surface water waves \cite{peter2006water,linton2007embedded,porter1999rayleigh,bennetts2017localisation}, thin elastic plates \cite{linton2002existence} and one-dimensional infinite array of point masses on an infinite, thin elastic plate \cite{chaplain2019rayleigh}, whispering gallery modes \cite{maling2016whispering}, lines of acoustic resonators inside a thick plate \cite{ward2022gapless}, and acoustic diffraction gratings in air \cite{chaplain2025acoustic}.

There have been several discussions of the relation of Rayleigh-Bloch waves to the trapped modes that appear due to rigid obstacles placed symmetrically in between parallel walls having either Neumann or Dirichlet conditions \cite{evans1998trapped,evans1999trapping,linton2007embedded,maniar1997wave}. The obvious difference is indeed the boundary condition at infinity being replaced by a boundary condition at a finite distance, hence the similarity remains limited to the perfectly evanescent RB waves whose dispersion lies below the sound cone.
The two-dimensional problem of acoustic scattering of an incident plane wave by a semi-inﬁnite array of either rigid or soft circular scatterers \cite{peter2006water,linton2007resonant,peter2007water,thompson2008new,bennetts2022rayleigh} reveals the existence of complex eigenfrequencies, corresponding to leaky, radiating wave solutions, whose dispersion extends inside the sound cone \cite{laudePRB2018}.
It is the purpose of this paper to show that leaky Rayleigh-Bloch waves can be guided inside the sound cone and that guided bound states in the continuum arise, as a result of the combination of symmetry and periodicity.
The concept of guided quasi-normal modes (QNMs) is proposed to obtain the complex dispersion relation of leaky Rayleigh-Bloch waves \cite{matsushima2024tracking} and the case of slow inclusions placed inside a fast medium is shown to be of particular interest.

\section{Velocity or impedance contrast?}

\begin{figure*}[tb]
\centering
\includegraphics[width=2\columnwidth]{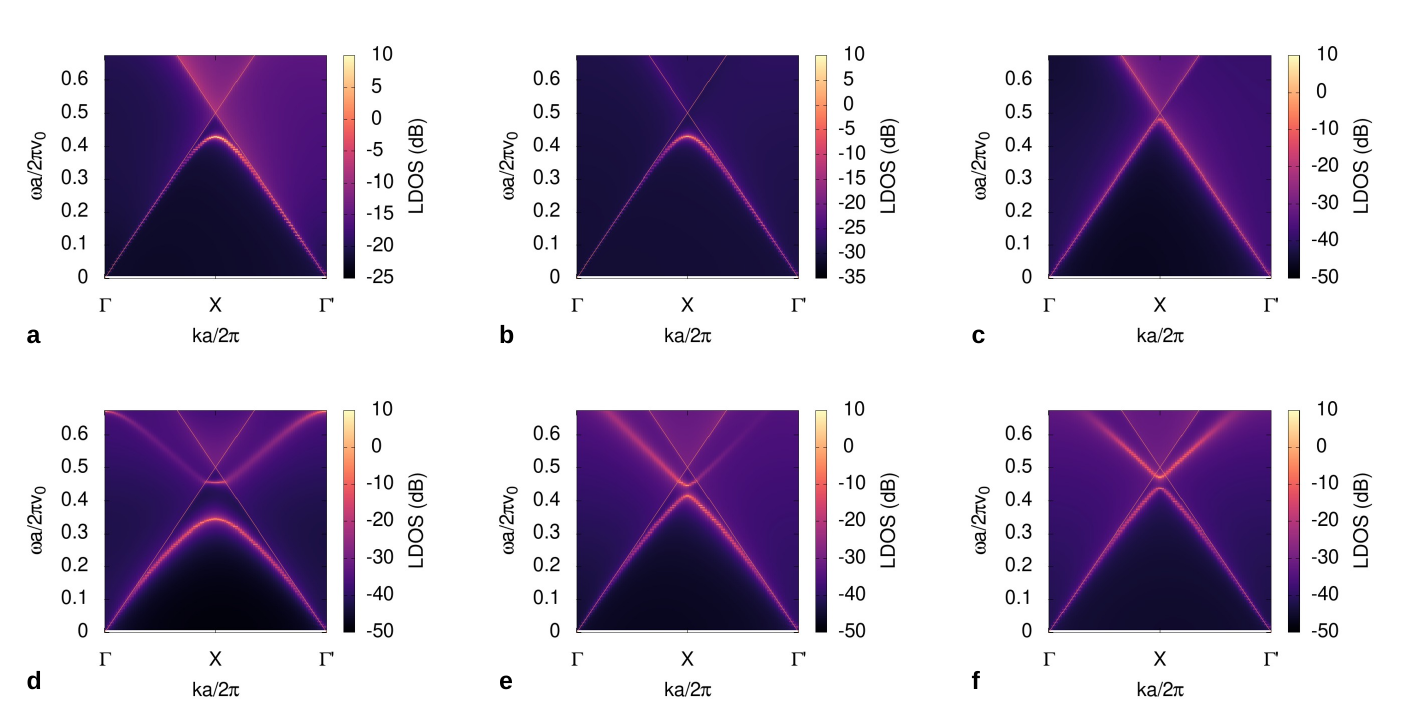}
\caption{Maps of the dispersion relation of Rayleigh-Bloch waves, computed using the resolvent band structure method \cite{laudePRB2018}. The color scale is for the local density of states (LDOS) estimated from the response to a stochastic source term. Panels are for (a) hollow inclusions satisfying a Neumann boundary condition along their edge, (b) inclusions with $\bar{\rho}=\bar{B}=100$ ($\bar{v}=1$ and $\bar{Z}=100$), (c) inclusions with $\bar{\rho}=\bar{B}=2$ ($\bar{v}=1$ and $\bar{Z}=2$), (d) inclusions with $\bar{\rho}=1$ and $\bar{B}=1/4$ ($\bar{v}=1/2$ and $\bar{Z}=1/2$), (e) inclusions with $\bar{\rho}=2$ and $\bar{B}=1/2$ ($\bar{v}=1/2$ and $\bar{Z}=1$), (f) inclusions with $\bar{\rho}=4$ and $\bar{B}=1$ ($\bar{v}=1/2$ and $\bar{Z}=2$).}
\label{fig2}
\end{figure*}

We consider in the following that the time-harmonic wave field satisfies a scalar Helmholtz equation, which includes the case of acoustic waves and of surface waves on water, for instance.
It is hence representative of the simplest wave models and could be extended to vector elastic waves in the future.
For concreteness of the discussion, we use acoustic notations in the following.
The wave equation is written
\begin{equation}
- (\nabla - \imath k \hat{x}) \cdot \left( \bar{\rho}^{-1} (\nabla - \imath k \hat{x}) p \right) - \omega^2 \bar{B}^{-1} p = \sigma
\label{eq1}
\end{equation}
with $\bar{\rho} = \rho / \rho_0$ the mass density relative to the host medium; similarly $\bar{B} = B / B_0$ is the relative elastic modulus.
Both dimensionless quantities are functions of space coordinates (Figure \ref{fig1}).
$\sigma$ is a body source term. 
The wave field has the dependence $p(x,y) \exp(\imath (\omega t - k x))$, with $p(x,y)$ periodic along axis $x$ and the primitive unit cell depicted in Figure \ref{fig1}.

Rayleigh-Bloch waves are most often considered in the case of hard-wall boundaries inside a fluid medium.
In this case a Neumann boundary condition along the edge of the inclusion applies (vanishing normal derivative of the wave field; $\partial_n p = 0$); alternatively the Dirichlet boundary condition can also be considered (vanishing wave field; $p=0$).
Figure \ref{fig2}a displays the dispersion relation computed using the resolvent formalism \cite{laudePRB2018}.
The diagram shows the local density of states (LDOS) estimated from the response to a stochastic source $\sigma$ in Eq. \eqref{eq1}, for every point $(k, \omega)$ in dispersion space.
In order to include scaling effects, the reduced wavenumber $\bar{k}=ka/(2\pi)$ and the reduced frequency $\bar{\omega}=\omega a/(2\pi v_0)$ are used, with $v_0=\sqrt{B_0/\rho_0}$ the velocity in the host medium.
As a note, in the different panels of Fig. \ref{fig2} there is a slight asymmetry in LDOS values with respect to the X point; this spurious imbalance in the response results numerically from the limited resolution of the finite element mesh used.
Below the sound cone, i.e. within the non-radiative region of dispersion space, there is a single band giving the dispersion relation for Rayleigh-Bloch waves, for hollow inclusions with a hard-wall boundary.
This is the usual solution considered in most papers on RB waves.
The single band is folded at the X point of the Brillouin zone because of periodicity.
A Bragg band gap opens for frequencies above it, but the upper band that closes this band gap is not apparent.
We show in the following that it actually locates inside the sound cone and is strongly subject to radiation loss, so that it does not leave a visible trace in the resolvent band structure.

We now consider that the inclusions are filled with another fluid instead of being hollow.
Heuristically, it can be understood that letting the dimensionless mass density of the inclusion tend to infinity results in a vanishing normal gradient of the wavefield along the inclusion boundary \footnote{A mathematical derivation can be obtained by applying the divergence theorem to a closed contour encircling locally the inclusion boundary.}.
Hence, we can consider a fictitious medium such that both $\bar{\rho}$ and $\bar{B}$ become very large in the same proportion.
As a result the dimensionless acoustic velocity $\bar{v} = \sqrt{\bar{B} / \bar{\rho}}=1$ remains constant whereas the dimensionless acoustic impedance $\bar{Z}=\sqrt {\bar{\rho}\bar{B}}$ increases in proportion.
It can be checked that the dispersion relation for $\bar{\rho}=100$ and $\bar{B}=100$ in Figure \ref{fig2}b is actually very close to the hollow inclusion case of Figure \ref{fig2}a.
It is further instructive to test a case with $\bar{\rho}$ = $\bar{B}$ not too large, in which case there is no velocity contrast but a moderate impedance contrast ($\bar{v}=1$ and $\bar{Z}=2$; see Fig. \ref{fig2}c).
There is still a single band, but the Bragg band gap tends to close toward the crossing point of right and left sound lines.
It is generally observed that the dispersion of Rayleigh-Bloch waves follows closely the sound lines when $\bar{v}=1$ and that the Bragg opening under the sound cone scales with the impedance contrast.

\begin{figure*}[t]
\centering
\includegraphics[height=45mm]{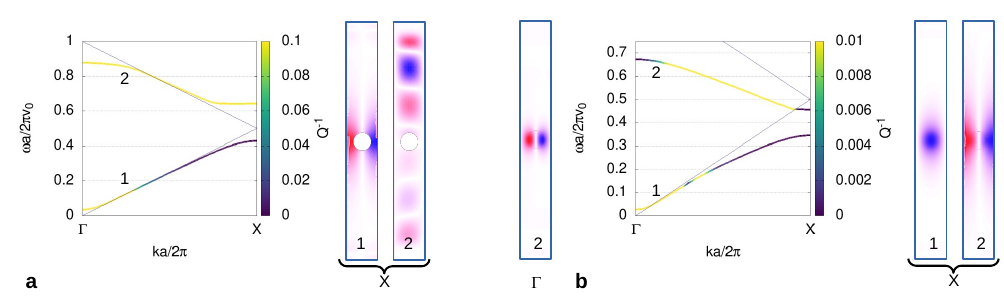}
\caption{
Complex dispersion relation for leaky Rayleigh-Bloch waves treated as guided quasi-normal modes.
The color scale is for the inverse of the quality factor.
(a) For hollow inclusions (same conditions as in Fig. \ref{fig2}a) the first band is guided but the second band, extending inside the sound cone, is always strongly leaky.
(b) For filled inclusions with $\bar{\rho}=1$ and $\bar{B}=1/4$ (same conditions as in Fig. \ref{fig2}d) the dispersion of the second band is lossless at both the X and the $\Gamma$ points.
The latter solution is a bound state in the continuum (BIC) whereas the former is a guided wave.
The real part of the pressure field of guided QNM solutions is shown within the primitive unit-cell at high-symmetry points.
}
\label{fig3} 
\end{figure*}

The situation changes if the inclusions are allowed to be slower than the host medium, in which case the dispersion of Rayleigh-Bloch waves shifts down in frequency as they localize more inside the inclusions (see Fig. \ref{fig2}d-f).
Significantly, the second band above the Bragg band gap now appears clearly and extends inside the sound cone, i.e. inside the radiative region of dispersion space.
The Bragg band gap itself exists even if the impedance contrast vanishes (case of Fig. \ref{fig2}e) and its opening is favored for $\bar{Z}<1$ (Fig. \ref{fig2}d) as compared to the inverse setting $\bar{Z}>1$ (Fig. \ref{fig2}f).

From the above observations, it can be concluded that the point in dispersion space around which the Bragg band gap opens for Rayleigh-Bloch waves can be moved down under the sound cone using slow inclusions ($\bar{v}<1$).
This results in the second or folded band to be much less leaky than in the usual case of hollow inclusions.
The opening of the Bragg band gap is then favored by decreasing the relative acoustic impedance of the inclusions ($\bar{Z}<1$).

\begin{figure*}[t]
\centering
\includegraphics[height=45mm]{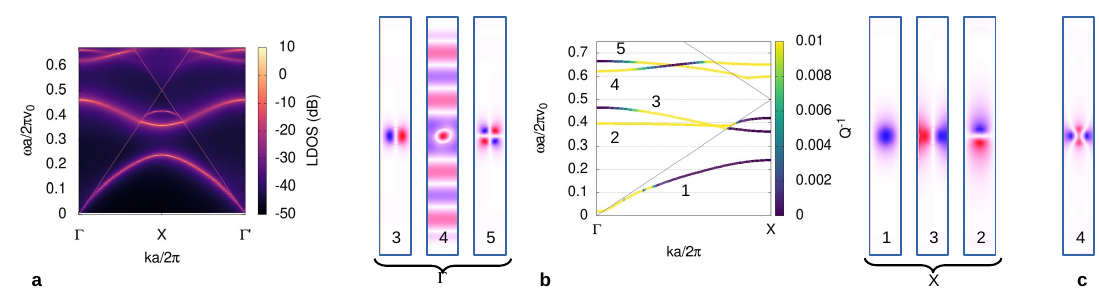}
\caption{
Dispersion relation of Rayleigh-Bloch waves for inclusions with $\bar{\rho}=1$ and $\bar{B}=1/9$ ($\bar{v}=1/3$ and $\bar{Z}=1/3$).
(a) Map of the dispersion relation computed using the resolvent band structure method \cite{laudePRB2018}. The color scale is for the local density of states (LDOS) estimated from the response to a stochastic source term.
(b) Complex dispersion relation for leaky Rayleigh-Bloch waves treated as guided quasi-normal modes.
The color scale is for the inverse of the quality factor.
The real part of the pressure field of guided QNM solutions is shown within the primitive unit-cell at high-symmetry points.
Examples of bound states in the continuum (BIC) are shown for bands 3 and 5 at the $\Gamma$ point.
(c) The Rayleigh-Bloch wave of band $4$ at $ka/(2\pi)\approx0.236$ inside the first Brillouin zone is also a BIC.
}
\label{fig4} 
\end{figure*}

\begin{figure*}[t]
\centering
\includegraphics[height=45mm]{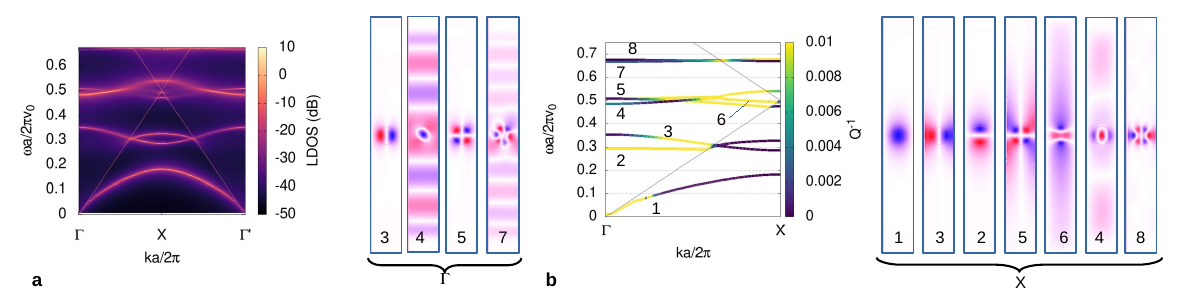}
\caption{
Dispersion relation of Rayleigh-Bloch waves for inclusions with $\bar{\rho}=1$ and $\bar{B}=1/16$ ($\bar{v}=1/4$ and $\bar{Z}=1/4$).
(a) Map of the dispersion relation computed using the resolvent band structure method \cite{laudePRB2018}. The color scale is for the local density of states (LDOS) estimated from the response to a stochastic source term.
(b) Complex dispersion relation for leaky Rayleigh-Bloch waves treated as guided quasi-normal modes.
The color scale is for the inverse of the quality factor.
The real part of the pressure field of guided QNM solutions is shown within the primitive unit-cell at high-symmetry points.
Examples of bound states in the continuum (BIC) are shown for bands 3 and 5 at the $\Gamma$ point, and band 8 at the X point.
}
\label{fig5} 
\end{figure*}

\section{Complex dispersion relation}
\label{sec3}

We now attempt to describe leaky RB waves that are partially guided along the chain of inclusions.
The concept of quasi-normal mode (QNM) \cite{chingRMP1998,wu2021nanoscale} is used to quantify the effect of radiation loss of a phononic resonator in an open medium \cite{laude2023quasinormal}.
For a QNM with a complex frequency $\omega$, in particular, the quality factor of the resonance is defined as $Q=\Re(\omega)/2\Im(\omega)$.
Bloch waves are the eigenfunctions of a periodic, lossless medium and the band structure is composed of their dispersion relation.
Here, we describe damped resonances in the resolvent band structure as quasi-normal Bloch waves, their complex eigenfrequencies defining complex bands $\omega_n(k)$.
The practical algorithm described in Ref. \cite{laude2023quasinormal} to obtain a single QNM is here modified to track the complex dispersion relation of guided QNMs as a function of the wavenumber.
Consider a discrete sequence $(k_i, i=0 \cdots m)$ sampling the $k$ axis.
Starting from a given point $(k_i,\omega_i)$ of dispersion space, the iteration converges fast toward the closest QNM at fixed $k_i$, yielding an estimate of the complex frequency $\omega_n(k_i)$ as well as of the field $p_n(x;k_i)$ of the QNM of interest.
If $i=0$ the starting frequency has to be guessed from the resolvent band structure and a stochastic source is chosen for initialization.
For step $i>0$, the starting frequency and the QNM candidate can be chosen as $\omega_n(k_{i-1})$ and $p_n(x;k_{i-1})$, and the iteration will converge to $\omega_{n}(k_{i})$ and $p_{n}(x;k_{i})$.
Repeating this elementary step, the complex band is easily and efficiently obtained, since each iteration requires the solution of only a few linear systems.

Considering the case of Rayleigh-Bloch waves for hollow inclusions in Fig. \ref{fig2}a, the complex band tracking algorithm readily produces the complex dispersion relation displayed in Fig. \ref{fig3}a.
The first band, that was visible in the resolvent band structure, is lossless (purely real) since it belongs to guided modes under the sound cone.
It is seen that the QNM tracking algorithm does not function very well at low frequencies and low wavenumbers, since the solution extends laterally very widely inside the PML.
The QNM can in this case not be clearly separated from PML eigenmodes and can be lost by the tracking algorithm.
Note that this limitation does not affect the computation of the resolvent band structure, anyway.
The second complex band has a low quality factor for all wavenumbers and could not be seen in the resolvent band structure, since its response is very low.
Moving the excitation frequency to the complex plane however makes it apparent.

Considering the slow inclusion case of Fig. \ref{fig2}d, the same procedure leads to the complex dispersion relation displayed in Fig. \ref{fig3}b.
The first complex band is similar to the one in Fig. \ref{fig3}a, though the wave field localizes on the inclusions rather than in between them.
Significantly, since the second complex band has moved down in frequency, it first appears as lossless after the folding at the X point of the Brillouin zone ($ka/(2\pi)=0.5$) but becomes lossy as it enters the sound cone.
Surprisingly, loss tends to $0$ ($Q \longrightarrow \infty$) at the $\Gamma$ point ($ka/(2\pi)=0$) for this second band.
At this point, the QN Bloch wave becomes a bound state in the continuum (BIC) \cite{stillinger1975bound, hsu2016bound,an2024multibranch}, since it it lossless although its dispersion lies inside the sound cone.
The BIC property here results from a combination of symmetry and periodicity, as the QN Bloch wave is a collective vibration state of the periodic, infinite chain of inclusions.
This is in contrast to band folding considered as the BIC generation mechanism \cite{wang2023brillouin}.

We next increase the material contrast in Fig. \ref{fig4}.
The frequency decrease of the dispersion of QN Bloch waves is stronger and a total of 5 bands is observed in the frequency range of interest, the first three of them being guided Rayleigh-Bloch waves, i.e. extending below the sound cone.
Each new band is clearly associated with a particular resonance of the inclusion and has a definite symmetry, in particular dictated by the azimuthal number $m$ of isolated QNMs discussed in Appendix \ref{appA}.
Band 3 ($m=1$) and band 5 ($m=2$) both hold a BIC at the $\Gamma$ point.
Surprisingly, another BIC occurs for band 4 for a $k$ value in between high symmetry points $\Gamma$ and X, at $ka/(2\pi)\approx0.236$.
As Fig. \ref{fig4}c shows, $m=2$ for this BIC also.

Increasing again the material contrast in Fig. \ref{fig5}, the overall trends are confirmed.
There are now 8 complex bands, with the additional appearance of a $m=3$ resonance leading to a BIC at both the $\Gamma$ and the X point (band 8).
Band 4 again holds a BIC for a $k$ value in between high symmetry points $\Gamma$ and X, at $ka/(2\pi)\approx0.172$, again with $m=2$.
The corresponding modal shape, not shown in Fig. \ref{fig5}, is very similar to the one in Fig. \ref{fig4}c.

\section{Conclusion}

The complex dispersion relation of Rayleigh-Bloch waves has been discussed for both the traditional case of hollow inclusions and the case of slow inclusions in a homogeneous propagation medium.
The introduction of slow inclusions, in particular, results in the appearance of localized resonances whose frequencies down-shift when velocity decreases, leading to the formation of additional Rayleigh-Bloch wave bands.
The key to the description of leaky Bloch waves that are partially guided along a periodic chain of inclusions is here the concept of phononic quasi-normal modes, that has been extended to include the case of guided waves. For a fixed real wavenumber, QNMs have a complex eigenfrequency that can be estimated by a search inside the complex dispersion plane. As a result, both the eigenmodes and the quality factor $Q$ of the associated resonance are obtained. Interestingly, for certain values of the wavenumber and as a result of its symmetry, a guided QNM can uncouple from bulk radiation modes, allowing to identify it as a bound state in the continuum (BIC).

The material system considered in the present derivation -- a fluid in a fluid -- may seem difficult to realize experimentally.
The reason for this choice was to handle the Helmholtz equation for scalar waves, one of the simplest among the class of wave equations. The results discussed here, however, already apply to the case of pure-shear out-of-plane elastic waves in solids, for instance, or of transverse-electric and transverse-magnetic electromagnetic waves.
It remains to extend the exposed method to vector Rayleigh-Bloch waves in solids and in three dimensions, but the concept incidentally shines a new light on previous experimental \cite{socie2013surface} and numerical \cite{al2016guidance} results for surface acoustic waves (SAW) guided along a chain of pillars on a substrate.

\section*{Acknowledgments}

This work was supported by the EIPHI Graduate School [grant number ANR-17-EURE-0002].
Finite element computations were performed with FreeFem$++$ \cite{freefem}.

\section*{Data availability statement}

Numerical data for resolvent band structures and codes to reproduce them are openly available at \cite{RBwaves}, together with an example code to obtain the QNMs of Fig. \ref{fig6}. Codes to reproduce complex dispersion relations are available upon reasonable request from the author.

\appendix
\section{QNM symmetry}
\label{appA}

The numerical observations made in Section \ref{sec3} hint at the importance of quasi-normal mode symmetry.
Rayleigh-Bloch waves, as as particular type of Bloch waves, are collective, periodic excitations.
As such they convey the symmetry of both the primitive unit-cell and of the periodic lattice.
In this appendix we examine the symmetry of the former, given here by the symmetry of the inclusion.

\begin{figure}[h]
\centering
\includegraphics[width=85mm]{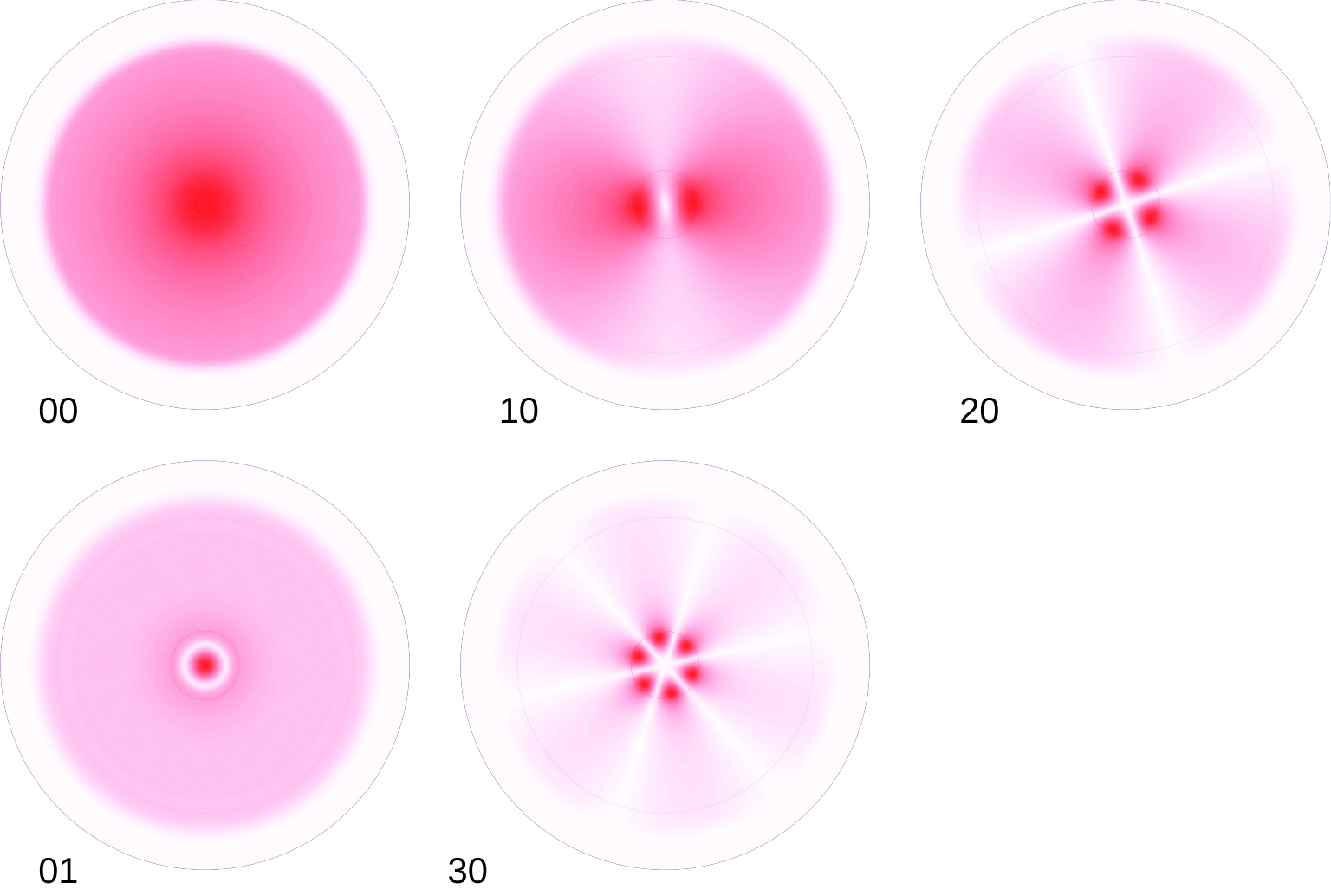}
\caption{
Quasi-normal modes of isolated slow inclusions with diameter $d=0.6a$. The labels refer to $m$ and $n$, the azimuthal and radial numbers. Quasi-normal modes are computed for the case $\bar{\rho}=1$, $\bar{B} = 1/16$ of Table \ref{tab1} and their modulus is displayed after normalization.
}
\label{fig6} 
\end{figure}

\begin{table}[th]
\caption{Properties of quasi-normal modes of isolated slow inclusions with diameter $d=0.6a$. $m$ and $n$ are the azimuthal and radial numbers, respectively.}
\label{tab1}
\begin{tabular}{l|lSSSSS}
\toprule
Contrast & $mn$ & {00} & {10} & {20} & {01} & {30} \\ \toprule
$\bar{\rho}=1$, $\bar{B} = 1/4$ & $\bar{\omega}$ & 0.39 & 0.59 & 0.96 & {$-$} & 1.26 \\
 & $Q$ & 1.5 & 3.4 & 4.7 & {$-$} & 10.5 \\ \midrule
$\bar{\rho}=1$, $\bar{B} = 1/9$ & $\bar{\omega}$ & 0.13 & 0.41 & 0.65 & 0.69 & 0.88 \\
 & $Q$ & 1.6 & 3.4 & 16.1 & 5.1 & 50.3 \\ \midrule
$\bar{\rho}=1$, $\bar{B} = 1/16$ & $\bar{\omega}$ & 0.11 & 0.30 & 0.49 & 0.53 & 0.67 \\
 & $Q$ & 2.0 & 6.7 & 30.1 & 8.5 & 171.7 \\ \bottomrule
\end{tabular}
\end{table}

Consider a single inclusion embedded in an infinite surrounding acoustic medium.
It support quasi-normal modes describing vibretions localized around the inclusion but radiating energy away from it.
This situation can be represented numerically using a PML surrounding completely the inclusion \cite{laude2023quasinormal}.
The first five QNMs that are found numerically in the frequency range of interest are displayed in Fig. \ref{fig6}.
Considering the central symmetry of the problem, the acoustic wave equation separates in polar coordinates $(r, \theta)$.
As a result, QNMs are indexed by an azimuthal number $m$ and a radial number $n$, so that $p_{mn}(r, \theta) = \exp(\imath m \theta) P_n(r)$.
They are doubly-degenerate if $m >0$.
The numerical QNM solutions thus converge to a superposition with azimuthal indices $\pm m$ in this case.
The result of the classification of QNMs is summarized in Table \ref{tab1}.
Quality factors are either low or moderate, because of radiation toward infinity, and improve with the azimuthal number.
The reduced QNM frequencies are in a clear correspondance with the Rayleigh-Bloch bands of figures \ref{fig3}b, \ref{fig4} and \ref{fig5}.
The BICs discussed in section \ref{sec3} are found for azimuthal numbers $m>0$.


%

\end{document}